\documentclass[12pt]{article}
\usepackage{times}
\usepackage{geometry}
\usepackage{xspace}
\usepackage[utf8]{inputenc}
\usepackage{enumitem,amssymb}
\usepackage{graphicx}
\usepackage{multicol}
\newlist{thematic}{itemize}{8}
\setlist[thematic]{label=$\square$}
\usepackage{pifont}

\def\aap{{\rm A\&A}}

\def\nat{{\rm Nature}}

\newcommand\ltsima{$\; \buildrel <\over\sim \;$}
\newcommand\simlt{\lower.5ex\hbox{\ltsima}}
\newcommand\gtsima{$\; \buildrel >\over\sim \;$}
\newcommand\simgt{\lower.5ex\hbox{\gtsima}}

\newcommand\mearth {{M_\oplus}}

\newcommand{\sigjit}[1]{
        \ensuremath{
                \ifthenelse{\equal{#1}{}}{\sigma_\mathrm{jit}}{\sigma_\mathrm{jit,#1}}}
        \xspace
}

\newcommand{\zfree}[1]{\ensuremath{\ifthenelse{\isempty{#1}}{z_{\mathrm{free}}^{*}}{z_{\mathrm{free,#1}}}}\xspace}

\newcommand\etal{et~al.}

\begin{document}
\begin{flushleft}
\huge
Astro2020 APC White Paper \linebreak

Community Involvement in the WFIRST Exoplanet Microlensing Survey \linebreak
\normalsize

  
\textbf{Principal Author:}

Name:	David P.\ Bennett
 \linebreak						
Institution:  NASA Goddard Space Flight Center and University of Maryland
 \linebreak
Email: david.bennett@nasa.gov
 \linebreak
Phone:  (301) 286-5473
 \linebreak
 
\textbf{Co-authors:} 
Rachel Akeson$^{1}$,
Thomas Barclay$^{2,3}$,
Jean-Phillipe Beaulieu$^{4,5}$,
Aparna Bhattacharya$^{2,3}$, 
Padi Boyd$^{2}$,
Valerio Bozza$^{10}$,
Geoffrey Bryden$^{7}$,
Sebastiano Calchi Novati$^{1}$,
Knicole Colon$^{2}$,
B.~Scott Gaudi$^{8}$,
Calen B.~Henderson$^{1}$,
Yuki Hirao$^{2,3,9}$,
Savannah Jacklin$^{33}$
Naoki Koshimoto$^{2,3,11}$,
Jessica Lu$^{12}$,
Matthew Penny$^{8}$,
Radek Poleski$^{8}$,
Elisa Quintana$^{2}$,
Cl\'ement Ranc$^{2}$,
Kailash C.~Sahu$^{13}$,
Rachel Street$^{14}$, 
Takahiro Sumi$^{9}$, 
Daisuke Suzuki$^{15}$,
Jennifer Yee$^{16}$
\vspace{0.3cm}
\begin{multicols}{2}
\noindent $^{1}$IPAC/Caltech \\
$^{2}$NASA Goddard Space Flight Center \\
$^{3}$University of Maryland \\
$^{4}$University of Tasmania, Australia \\
$^{5}$Institut d'Astrophysique de Paris, France \\
$^{6}$University of Salerno, Italy \\
$^{7}$Jet Propulsion Laboratory \\
$^{8}$Ohio State University \\
$^{9}$Osaka University, Japan \\
$^{10}$Vanderbilt University \\
$^{11}$University of Tokyo \\
$^{12}$University of California, Berkeley \\
$^{13}$Space Telescope Science Institute \\
$^{14}$Las Cumbres Observatory \\
$^{15}$ISAS, JAXA, Japan \\
$^{16}$Harvard-Smithsonian CfA \\
\end{multicols}

\textbf{Abstract:}
WFIRST is NASA's first flagship mission with pre-defined core science programs to study dark energy
and perform a statistical census of wide orbit exoplanets with a gravitational microlensing survey.
Together, these programs are expected to use more than half of the prime mission observing time. 
Previously, only smaller, PI-led missions have had core programs that used such a large fraction 
of the observing time, and in many cases, the data from these PI-led missions was reserved for the
PI's science team for a proprietary period that allowed the PI's team to make most of the major 
discoveries from the data. Such a procedure is not appropriate for a flagship mission, which should
provide science opportunities to the entire astronomy community. For this reason, there will be no
proprietary period for WFIRST data, but we argue that a larger effort to make WFIRST science 
accessible to the astronomy community is needed. We propose a plan to enhance community
involvement in the WFIRST exoplanet microlensing survey in two different ways. First, we propose
a set of high level data products that will enable astronomers without detailed microlensing expertise
access to the statistical implications of the WFIRST exoplanet microlensing survey data. And second,
we propose the formation of a WFIRST Exoplanet Microlensing Community Science Team that will open
up participation in the development of the WFIRST exoplanet microlensing survey to the general 
astronomy community in collaboration for the NASA selected science team, which will have the
responsibility to provide most of the high level data products. This community science team will be open to 
volunteers, but members should also have the opportunity to apply for funding.

\end{flushleft}
\pagebreak
\section{The WFIRST Exoplanet Microlensing Survey}
\vspace{-0.3cm}

The science of the WFIRST Exoplanet Microlensing survey was submitted to the Astro2010 decadal
survey in the form of a smaller mission concept, known as the Microlensing Planet Finder (MPF) 
(Bennett et al.\ 2004, 2010). This mission concept was combined with two other 
wide field infrared telescope 
mission concepts, JDEM-$\Omega$ (Gehrels 2010) and NIRSS (Stern et al.\ 2010), to form
the WFIRST mission. The exoplanet microlensing survey of the MPF mission and the dark energy
surveys of the JDEM-$\Omega$ mission were called out as high priority science goals for WFIRST,
with the expectation that the majority of observing time would be devoted to these two programs.
The prominence of large surveys in astronomy has been a growing trend over the last several decades,
leading up to the Astro2010 decadal survey that selected two survey telescopes, WFIRST and LSST,
as the top large space and ground-based programs. 

\begin{figure}[h]
 \begin{minipage}[t]{0.73\textwidth}
  \includegraphics[width=\textwidth]{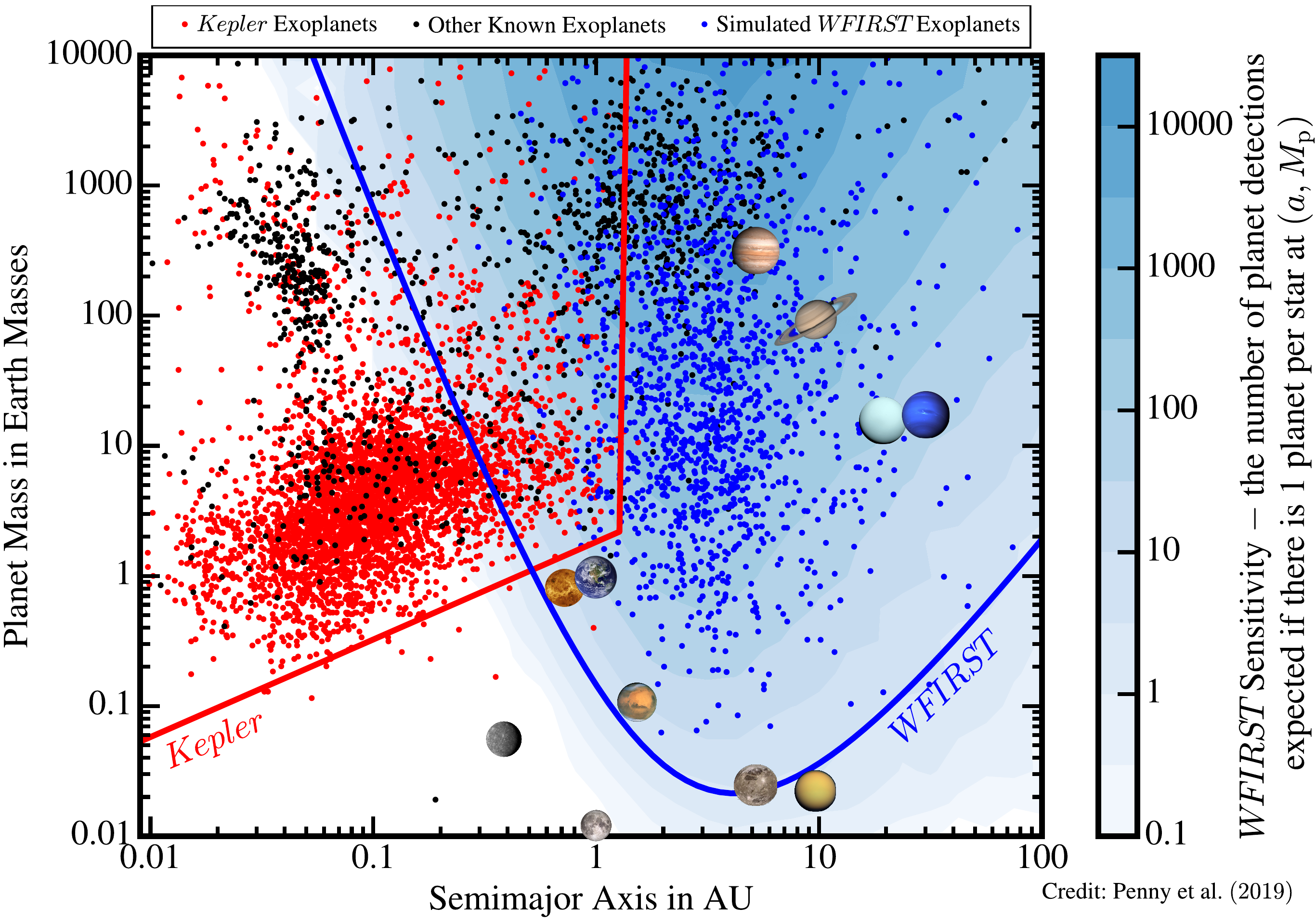}
 \end{minipage}\hfill
 \begin{minipage}[b]{0.26\textwidth}
\caption{Comparison of planet mass and semimajor axis for planets candidates found by Kepler (red), 
planet discoveries by other current methods (black) and simulated planets from WFIRST (blue). 
The pictures show the locations of solar system
planets and moons (as if they are planets).}
 \end{minipage}
\end{figure}

Figure 1 shows the mass and semi-major axis distribution of the expected WFIRST exoplanet discoveries with 
Kepler's planet candidates and planets found by other methods (Penny \etal\ 2019).
WFIRST will complete the statistical census of exoplanets started by Kepler with 
high sensitivity to low-mass planets
in wide orbits, ranging from the habitable zone of FGK stars to infinity, i.e.\ unbound planets
(Mr{\'o}z \etal\ 2019).
As Figure~1 indicates, WFIRST is sensitive to analogs of all the planets in our Solar System,
except for Mercury, and has sensitivity extending down to planets below the mass of Mars 
($\simlt 0.1\mearth$). This is more than 2 orders of magnitude lower in mass than
other methods.

\begin{figure}[h]
 \begin{minipage}[t]{0.65\textwidth}
  \includegraphics[width=\textwidth]{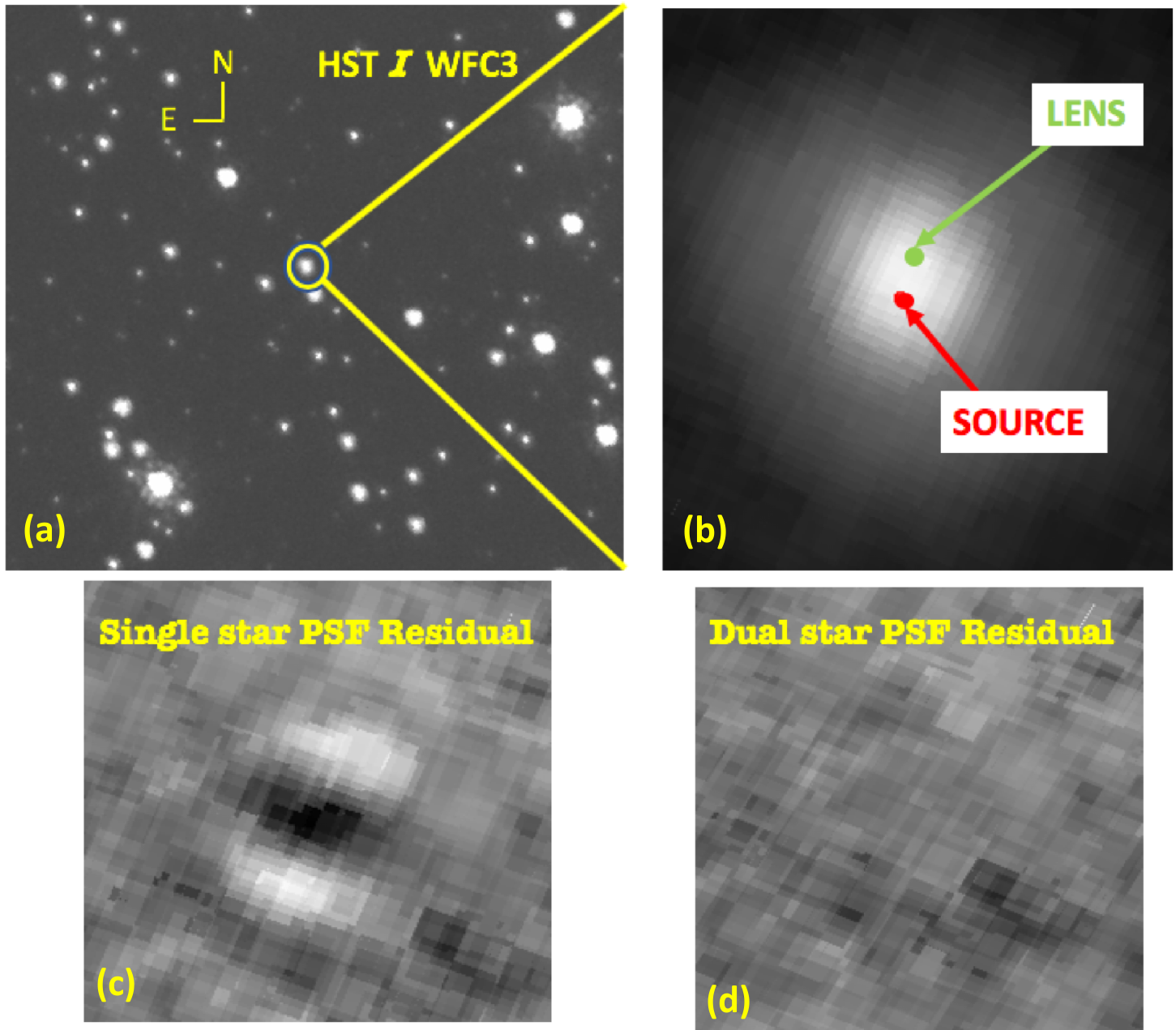}
 \end{minipage}\hfill
 \begin{minipage}[b]{0.32\textwidth}
\caption{Coadded HST $I$-band image, (a), and a close-up image of the blended
lens+source images, (b), indicating the inferred positions of the OGLE-2012-BLG-0950
source and lens stars. The residual from the best fit single star fit indicates that two stars contribute
to the blended image (c), while the dual star (lens+source) PSF fit leaves a clean residual (d).}
 \end{minipage}
\end{figure}

While microlensing light curves themselves usually yield only the star-planet mass ratio, $q$, WFIRST's 
space-based imaging will yield host star and planet masses for the majority of planets
discovered using a combination of microlensing light curve constraints and direct observations
of the host star (Bennett \etal\ 2007). Figure~2 shows an example of
host star detection with HST for a planetary microlensing event observed from the ground (Bhattacharya \etal\ 2018)
that enabled both the mass and distance of the host star to be determined.
Microlensing is also sensitive to
planets in binary systems (Gould \etal\ 2013; Bennett \etal\ 2016), and WFIRST will be sensitive
to exomoons in systems similar to the Earth-moon system (Bennett \& Rhie 2002).

\vspace{-0.3cm}
\section{WFIRST Microlensing Survey High Level Data Products}
\vspace{-0.3cm}

The study of exoplanets has only developed into a major subfield of astronomy in the past 
quarter century due to the technical challenges of exoplanet detection. However,
microlensing is conceptually the most obscure of the major planet detection methods. Very few
astronomers have the background to understand the relationship between planetary microlensing light curves
and the associated planetary system properties. So, the astronomy community 
could really benefit from data products that handle all these microlensing subtleties. Therefore, the
WFIRST Project plans to provide a range of high level data products 
that will handle these microlensing subtleties in order to enable broader participation in the 
high profile exoplanet microlensing science that was called out by the Astro2010 decadal survey.
These high level data products include
the following image and microlensing light curve data products:
\begin{enumerate}
\item Light curve photometry and astrometry of all the stars identified in the WFIRST microlensing 
survey fields. The photometry will include corrections for all known systematic effects, including
the effect of the proper motion of neighboring stars and detector artifacts. 
The astrometry will include the relative astrometry
of unresolved lens and source stars as illustrated in Figure 2. The photometry will be updated daily
when the microlensing survey is in progress, and new improved photometry and astrometry will be
released at the end of each microlensing observing season.
\item Light curve models for all single, binary, and most triple lens microlensing events will be provided
at the end of each microlensing season, and most of these will be provided in real time as the data
come in during the microlensing observing seasons. The end of season releases will include the higher
order microlensing effects, such as microlensing parallax and orbital motion of the lens and source stars. 
\end{enumerate}

The community's experience with Kepler (Thompson et al. 2018) has shown that it is critical to 
calculate detection efficiencies and reliability in order to make robust conclusions regarding the statistical 
properties of exoplanets.  Within the field of microlensing, the importance of detection efficiencies 
(Rhie et al. 2000; Gaudi et al. 2002) was known before the first microlensed exoplanet was 
discovered (Bond et al. 2004). The reliability of planet detections is considered
to be less of a problem for microlensing than for transits because a smaller fraction of planetary signals are 
expected to be close to the S/N threshold (Gould, et al. 2004). However, in some cases it is possible
for the microlensing of a faint companion to the source star to mimic a planetary signal, although
high quality light curve data can usually rule out such scenarios (Gaudi 1998; Beaulieu et al. 2006).
There are also cases with degeneracies between star plus planet and binary star lens system models
for some events, which can be dealt with in a Bayesian analysis if the posterior distributions
for each of the degenerate models is known (Suzuki et al.\ 2016).
These considerations call for additional data products to be produced:
\begin{enumerate}
\item[3.] Detection efficiencies for every detected microlensing event, including those without detected 
planets, as a function of mass ratio, separation, source radius crossing time, and host star mass, when
it is available.
\item[4.] Posterior distributions from Markov Chain calculations for the parameters of each event. 
For events with degenerate solutions, including degeneracies between star plus planet lensing events
and binary source events, the posterior distributions will include the relative probabilities 
of the different solutions.
\item[5.] Reliability estimates for apparent planetary events with low S/N planetary signals, if these are included
the catalog of microlensing events.
\end{enumerate}

The current WFIRST plans call for high level data products 1 and 3 to be provided by the WFIRST Science Support
Center at IPAC, with algorithms provided by the WFIRST Microlensing Implementation Science Team (MIST), which
will also have the responsibility for providing the remaining high level data products. Note that the
MIST has yet to be selected. It is expected to be selected in
late 2021 after the term of the current Microlensing Science Investigation Team (MicroSIT) has expired.

\vspace{-0.3cm}
\section{WFIRST Microlensing Community Science Team}
\vspace{-0.3cm}

Under current plans, the WFIRST MIST would work 
to prepare the WFIRST Exoplanet Microlensing survey plans and the software needed to provide
the high level data products described above. It would then provide data products 2, 4, and 5 described
above. The MIST is expected to work 
from its selection, now planned for late 2021, through the WFIRST data collection phase,
from 2026 to 2030. Under the current plans, there would be few opportunities for non-MIST members 
to contribute to the development of the microlensing survey despite the 5-year interval between the
MIST selection and the beginning of the microlensing survey.
There would also be little
chance for outsiders to contribute to the microlensing survey with ideas that might enhance the
microlensing science itself or additional science that might come from the microlensing survey
(Gaudi et al.\ 2019).

We propose to address this problem by creating a WFIRST Microlensing Survey Community Science
Team (CST) analogous to the LSST Science Collaborations and the TESS science 
working groups. The goals of this WFIRST Microlensing CST would be to advance both the exoplanet
microlensing science and the auxiliary science to be produced by the microlensing survey.
The CST would be largely be open to volunteers, with  low barriers to entry for scientists 
external to microlensing. The CST
would include members who intend to contribute the exoplanet microlensing science, as
well as members who would like to use the microlensing science data or software for other 
science goals.

CST members interested in the exoplanet microlensing survey might contribute
to the photometry/astrometry or light curve modeling software being developed by the MIST.
This work would be done in collaboration with the MIST. That way,
a CST member could improve only one part of the software while using the 
rest of the software being developed by the MIST. The current MicroSIT
is working to provide a number of software tools to allow newcomers to 
analyze microlensing events, and the MOA Collaboration has released its 9-year microlensing light 
curve database to the NASA Exoplanet Archive as a part of Japan's contribution to WFIRST. 
(This MOA data sets includes the events used for the largest statistical study of microlens exoplanets
to date (Suzuki et al.\ 2016).) The 
MIST should extend this effort and welcome CST
members who want to learn how to work with and analyze exoplanet microlensing data.

The CST should also welcome members who want to pursue other science
goals using the microlensing survey data. This would include people working
on variable stars or other microlensing science, such as the search for black hole microlensing
events, as well as a variety of other science (Gaudi et al.\ 2019). These CST 
members might write additions to the microlensing survey photometry pipeline to 
characterize different types of variable stars or to identify and track Kuiper Belt objects or
asteroids as they orbit through the WFIRST fields. They could also write software to provide
high level data products to address additional science goals. Other CST members might
suggest modifications of the microlensing survey to enhance non-exoplanet microlensing
science goals. If these modifications have only a small impact on the exoplanet microlensing
science, the MIST might decide to accept such a modification. If there are several, incompatible
modifications suggested, sub-groups of the CST might propose to the WFIRST General Observer
(GO) Time Allocation Committee (TAC) to decide which modification to implement. For proposed
modifications that might have a serious impact on the microlensing survey sensitivity, the 
CST could work with the MIST to determine this impact and how much observing time might
be needed to compensate for the reduced exoplanet microlensing sensitivity that the modification
would entail. This would allow the WFIRST GO TAC to consider such a program without having
to consider a modification of the survey science.

While the CST would be open to unpaid volunteers, it will need to have associated funding 
opportunities to be fulfill its promise. The MIST will need to have funding to support the CST,
but members of the CST will also need access to funding. Typically, we might expect 
CST members to join as volunteers and then to propose for funding once they have identified a good 
proposal topic through their involvement with the CST.


We suspect that the other WFIRST surveys, the High Latitude Survey and the Supernova Survey,
might also benefit from similar program, but we leave it those surveys to try make plans for
CSTs for these other surveys.

\pagebreak
\noindent\textbf{References}\\
Beaulieu, J.-P., Bennett, D.~P., Fouqu{\'e}, P., et al.\ 2006, \nat, 439, 437 \\
Bennett, D.~P., Anderson, J., \& Gaudi, B.~S.\ 2007, ApJ, 660, 781 \\
Bennett, D.~P., Anderson, J., Beaulieu, J.-P., et al.\ 2010, arXiv:1012.4486 \\
Bennett, D.~P., Bond, I., Cheng, E., et al.\ 2004, SPIE, 5487, 1453 \\
Bennett, D.~P., \& Rhie, S.~H.\ 2002, ApJ, 574, 985 \\
Bennett, D.~P., Rhie, S.~H., Udalski, A., et al.\ 2016, AJ, 152, 125 \\
Bhattacharya, A., Beaulieu, J.-P., Bennett, D.~P., et al.\ 2018, AJ, 156, 289 \\
Bond, I.~A., Udalski, A., Jaroszy{\'n}ski, M., et al.\ 2004, ApJL, 606, L155 \\
Gaudi, B.~S., 1998, ApJ, 506, 533  \\
Gaudi, B.~S., Akeson, R, Anderson, J., et al.\ 2019, Astro2020 science white paper, arXiv:1903.08986 \\
Gaudi, B.~S., Albrow, M.~D., An, J., et al.\ 2002, ApJ, 566, 463  \\
Gould, A., Gaudi, B.~S., \& Han, C., 2004, arXiv:astro-ph/0405217 \\
Gould, A., Udalski, A., Shin, I.-G., et al.\ 2014, Science, 345, 46 \\
Mr{\'o}z, P., Udalski, A., et al.\ 2019, \aap, 622, A201\\
Penny, M.~T., Gaudi, B.~S., Kerins, E., et al.\ 2019, ApJS, 241, 3 \\
Rhie, S.~H., Bennett, D.~P., Becker, A.~C., et al.\ 2000, ApJ, 533, 378 \\
Stern, D., Bartlett, J.~D., Brodwin, M., et al.\ 2010 arXiv:1008.3563 \\
Suzuki, D., Bennett, D.~P., Sumi, T., \etal\ 2016, ApJ, 833, 135 (S16) \\
Thompson, S.~E., Coughlin, J.~L., Hoffman, K., et al.\ 2018, ApJS, 235, 38 \\
Yee, J.~C.,  Akeson, R, Anderson, J., \etal\ 2019, Astro2020 science white paper, arXiv:1903.08219 \\

\end{document}